\documentclass{bmcart}
\pdfoutput=1

\usepackage{amsthm,amsmath}

\usepackage{graphicx}

\startlocaldefs

\usepackage{acronym}
\usepackage{amssymb}
\usepackage[utf8x]{inputenc}
\usepackage[ruled,vlined,linesnumbered]{algorithm2e}
\usepackage{nicefrac}

\acrodef{osa}[OSA]{opportunistic spectrum access}
\acrodef{pu}[PU]{primary user}
\acrodef{su}[SU]{secondary user}
\acrodef{tdt}[TDT]{time-domain test}
\acrodef{ca}[CA]{cyclic autocorrelation}
\acrodef{sc}[SC]{spectral correlation}
\acrodef{cs}[CS]{compressive sampling}
\acrodef{omp}[OMP]{orthogonal matching pursuit}
\acrodef{mwc}[MWC]{modulated wideband converter}
\acrodef{scm}[SCM]{slot comparison method}
\acrodef{sm}[SM]{symmetry method}
\acrodef{dft}[DFT]{discrete Fourier transform}
\acrodef{idft}[IDFT]{inverse discrete Fourier transform}
\acrodef{dc}[DC]{direct current}
\acrodef{snr}[SNR]{signal to noise ratio}
\acrodef{somp}[SOMP]{simultaneous orthogonal matching pursuit}
\acrodef{cfar}[CFAR]{constant false alarm rate}
\acrodef{pdf}[PDF]{probability density function}
\acrodef{awgn}[AWGN]{additive white Gaussian noise}
\acrodef{mse}[MSE]{mean squared error}
\newcommand{\sober}{SOber} %
\newcommand{\hades}{Dice} %

\usepackage{cleveref} 
\crefformat{equation}{(#1)} 
\Crefformat{equation}{(#1)} 
\let\ref\Cref    

\newcommand{\ie}[0]{, i.\,e., }
\newcommand{\eg}[0]{, e.\,g., }

\newcommand{\ud}{\ensuremath{\,\mathrm{d}}} 
\newcommand{\vecop}[1]{\ensuremath{\operatorname{vec}\left\{ #1 \right\}}}
\newcommand{\T}{^\text{T}} 
\renewcommand{\H}{^\text{H}} 
\newcommand{\liminfty}[1]{ \underset{#1\rightarrow \infty}{\operatorname{lim}} } 
\newcommand{\convdist}{\ensuremath{\overset{\text{D}}{=}}} 
\newcommand{\realpart}[1]{\ensuremath{ \mathfrak{Re}\left\{ #1 \right\} }} 
\newcommand{\imagpart}[1]{\ensuremath{ \mathfrak{Im}\left\{ #1 \right\} }} 
\newcommand{\lpnorm}[2]{\left\| #1 \right\|_{\ell_{#2}}} 
\newcommand{\argmax}[2]{ \underset{#2}{\operatorname{argmax}}\left\{ #1 \right\} } 
\newcommand{\argmin}[2]{ \underset{#2}{\operatorname{argmin}}\left\{ #1 \right\} } 
\newcommand{\supp}[1]{ \operatorname{supp}\left(#1\right) } 
\newcommand{\integers}{\ensuremath{\mathbb{Z}}} 
\newcommand{\realnum}{\ensuremath{\mathbb{R}}}  
\newcommand{\complexnum}{\ensuremath{\mathbb{C}}}  
\renewcommand{\vec}[1]{\ensuremath{\boldsymbol{\mathbf{\MakeLowercase{#1}}}}}
\newcommand{\mat}[1]{\ensuremath{\boldsymbol{\mathbf{\MakeUppercase{#1}}}}}
\newcommand{\identitymat}[1]{\ensuremath{\mat{I}_{#1}}}
\newcommand{\hypzero}{\ensuremath{\mathcal{H}_0}} 
\newcommand{\hypone}{\ensuremath{\mathcal{H}_1}} 
\newcommand{\normal}[2]{\ensuremath{\mathcal{N}(#1, #2)}}

\usepackage{xifthen}

\newcommand{\shortminus}{\scalebox{0.5}[1.0]{\( - \)}}
\newcommand{\obssig}[1]{ \ifthenelse {\isempty{#1}} {x} {x(#1)} } 
\newcommand{\obssigconj}[1]{ \ifthenelse {\isempty{#1}} {x^\ast} {x^\ast(#1)} } 
\newcommand{\dobssig}[1]{\vec{x}_{#1}} 
\newcommand{\dobssigconj}[1]{\vec{x}^\ast_{#1}} 
\newcommand{\propsig}[1]{s'(#1)} 
\newcommand{\recnoise}[1]{\eta(#1)} 
\newcommand{\transsig}[1]{ \ifthenelse {\isempty{#1}} {s} {s(#1)} } 
\newcommand{\symb}[0]{c} 
\newcommand{\pulseshape}[1]{ \ifthenelse {\isempty{#1}} {p} {p\left(#1\right)} } 
\newcommand{\conjpulseshape}[1]{p^\ast\left(#1\right)} 
\newcommand{\transca}[2]{ \ifthenelse {\isempty{#1}} {R_{\transsig{}, \symblen{}}^\cycfreq(\delay)} {R_{\transsig{}, \symblen{}}^{#1}(#2)} } 
\newcommand{\randsigca}[2]{ \ifthenelse {\isempty{#2}} {R_{\symb{}}(#1)} {R_{\symb{}}^{#2}(#1)}} 
\newcommand{\pulseshapeca}[1]{r_{\pulseshape{}}^{\cycfreq}(#1)} 
\newcommand{\sigpow}[0]{\sigma^2_{\symb}} 
\newcommand{\cycfreq}[0]{\alpha} 
\newcommand{\dcycfreq}[0]{a} 
\newcommand{\atest}{\ensuremath{a_0}} 
\newcommand{\cavec}[1]{ \ifthenelse {\isempty{#1}} {\hat{\vec{r}}_{\obssig{}}^{\ddelay}} {\hat{\vec{r}}_{\obssig{}}^{#1}}  } 
\newcommand{\analcavec}[3]{ \ifthenelse {\isempty{#2}} {\vec{r}_{\transsig{}, #1}^{#3}} {\vec{r}_{\transsig{}, #1}^{#3}[#2]} } 
\newcommand{\delay}[0]{\tau} 
\newcommand{\ddelay}[0]{\nu} 
\newcommand{\ndelays}[0]{{n_\ddelay}} 
\newcommand{\ca}[2]{ \ifthenelse {\isempty{#1}} {R_{\obssig{}}^\cycfreq(\delay)} {R_{\obssig{}}^{#1}(#2)} } 
\newcommand{\dca}[3]{ \ifthenelse {\isempty{#1}} {{R'}_{\transsig{}, \symblen{}}^\cycfreq(\ddelay\smplper)} {{R'}_{\transsig{}, #3}^{#1}(#2)} } 
\newcommand{\caest}[1]{\hat{R}_{\obssig{}, \timezero}^{#1}(\ddelay)} 
\newcommand{\obsdur}[0]{T} 
\newcommand{\smplper}[0]{T_e} 
\newcommand{\symblen}[0]{T_s} 
\newcommand{\timezero}[0]{t_0} 
\newcommand{\blocksize}[0]{N} 
\newcommand{\navailsmpls}[0]{m} 
\newcommand{\delprodvec}[2]{\vec{y}_{#2}^{#1}} 
\newcommand{\delprodmat}[1]{\mat{Y}_{#1}} 
\newcommand{\dftmat}{\mat{F}} 
\newcommand{\idftmat}{\mat{F}^{-1}} 
\newcommand{\nsym}[0]{n_s} 
\newcommand{\pd}{P_\text{d}} 
\newcommand{\pfa}{P_\text{fa}} 
\newcommand{\camat}[0]{\hat{\bf R}_x} 
\newcommand{\testvec}{\ensuremath{\hat{\bf r}_{xx^\ast}}} 
\newcommand{\puretestvec}{\ensuremath{{\bf r}_{xx^\ast}}} 
\newcommand{\estnoise}{\ensuremath{\boldsymbol{\epsilon}_{xx^\ast}}} 
\newcommand{\covmat}{\ensuremath{ {\bf \Sigma}_{xx^\ast} }}
\newcommand{\estcovmat}{\ensuremath{ \hat{\bf \Sigma}_{xx^\ast} }}
\newcommand{\qmat}{\mat{Q}}
\renewcommand{\sc}[2]{S_{ \delprodvec{\ddelay_m}{\blocksize} \delprodvec{\ddelay_n}{\blocksize} }(#1, #2)} 
\newcommand{\scconj}[2]{S^\ast_{ \delprodvec{\ddelay_m}{\blocksize} \delprodvec{\ddelay_n}{\blocksize} }(#1, #2)} 
\newcommand{\estsc}[2]{\hat{S}_{ \delprodvec{\ddelay_m}{\blocksize} \delprodvec{\ddelay_n}{\blocksize} }(#1, #2)} 
\newcommand{\estscconj}[2]{\hat{S}^\ast_{ \delprodvec{\ddelay_m}{\blocksize} \delprodvec{\ddelay_n}{\blocksize} }(#1, #2)} 
\newcommand{\teststat}[1]{ \ifthenelse {\isempty{#1}} {\mathcal{T}} {\mathcal{T}_{#1}} } 
\newcommand{\samplmat}[0]{\mat{M}} 
\newcommand{\atomsmat}[0]{\mat{A}} 
\newcommand{\corrmat}[0]{\mat{C}} 
\newcommand{\ringcorrmat}[0]{\mathring{\mat{C}}} 
\newcommand{\corrvec}[1]{\vec{c}_{#1}} 
\newcommand{\ringcorrvec}[1]{\mathring{\vec{c}}_{#1}} 
\newcommand{\symdict}[1]{ \ifthenelse {\isempty{#1}} {\mat{D}_\text{sym}} {\mat{D}^{(#1)}_\text{sym}} } 
\newcommand{\symdictf}[1]{ \ifthenelse {\isempty{#1}} {\mathring{\mat{D}}_\text{sym}} {\mathring{\mat{D}}^{(#1)}_\text{sym}} } 
\newcommand{\asydictf}[1]{ \ifthenelse {\isempty{#1}} {\mathring{\mat{D}}_\text{sym}} {\mathring{\mat{D}}^{(#1)}_{\text{asy},l}} } 
\newcommand{\structdictf}[1]{ \ifthenelse {\isempty{#1}} {\mathring{\mat{D}}} {\mathring{\mat{D}}^{(#1)}} } 
\newcommand{\niter}{\ensuremath{n_\text{iter}}}
\newcommand{\half}{\frac{1}{2}}
\newcommand{\anhalf}{\frac{\dcycfreq}{2\blocksize}}
\newcommand{\consecfrac}[0]{\beta}

\endlocaldefs

\begin{document}

\begin{frontmatter}

\begin{fmbox}
\dochead{Research}


\title{Compressive Cyclostationary Spectrum Sensing with a Constant False Alarm Rate}


\author[
addressref={ti},
corref={ti},
email={bollig@ti.rwth-aachen.de}
]{\inits{AB}\fnm{Andreas} \snm{Bollig}}
\author[
addressref={ti},
email={arts@ti.rwth-aachen.de}
]{\inits{MA}\fnm{Martijn} \snm{Arts}}
\author[
addressref={il},
email={anastasia.lavrenko@tu-ilmenau.de}
]{\inits{AL}\fnm{Anastasia} \snm{Lavrenko}}
\author[
addressref={ti},
email={mathar@ti.rwth-aachen.de}
]{\inits{RM}\fnm{Rudolf} \snm{Mathar}}

\address[id=ti]{
	\orgname{Institute for Theoretical Information Technology, RWTH Aachen University},
	\street{Kopernikusstraße 16},
	\postcode{52074},
	\city{Aachen},
	\cny{Germany}
	}
\address[id=il]{
	\orgname{Institute for Information Technology, TU Ilmenau},
	\street{Helmholtzbau},
	\postcode{98684},
	\city{Ilmenau},
	\cny{Germany}
}


\begin{artnotes}
\end{artnotes}

\end{fmbox}


\begin{abstractbox}

\begin{abstract} 
Spectrum sensing is a crucial component of opportunistic spectrum access schemes, which aim at improving spectrum utilization by allowing for the reuse of idle licensed spectrum. 
Sensing a spectral band before using it makes sure the legitimate users are not disturbed. Since information about these users' signals is not necessarily available, the sensor should be able to conduct so-called \emph{blind} spectrum sensing.
Historically, this has not been a feature of cyclostationarity-based algorithms.
Indeed, in many application scenarios the information required for traditional cyclostationarity detection might not be available, hindering its practical applicability. 
In this work we propose two new cyclostationary spectrum sensing algorithms that make use of the inherent sparsity of the cyclic autocorrelation to make blind operation possible.
Along with utilizing sparse recovery methods for estimating the cyclic autocorrelation, we take further advantage of its structure by introducing joint sparsity as well as general structure dictionaries into the recovery process. Furthermore, we extend a statistical test for cyclostationarity to accommodate \emph{sparse} cyclic spectra.
Our numerical results demonstrate that the new methods achieve a near constant false alarm rate behavior in contrast to earlier approaches from the literature.
%
\end{abstract}




\end{abstractbox}
%

\end{frontmatter}



\section{Introduction}
\label{sec:intro}

The scarcity of radio spectrum constitutes a major roadblock to current and future innovation in wireless communications. 
To alleviate this problem, it has been proposed to make spectral resources, which are currently underutilized, available for reuse under a paradigm that goes by the name of \ac{osa} \cite{zhao_survey_2007}. 
Spectrum sensing is one of its core technologies.
It allows an unlicensed transceiver, a so called \ac{su}, to access a licensed spectral band without interfering with the owner of the band's license, the so called \ac{pu}. The fundamental task in spectrum sensing is to decide between two hypotheses, the first of which states that the spectral band under investigation is free (\hypzero), while the second asserts that it is occupied (\hypone). Considering the baseband signal $\obssig{t}$ observed at a secondary system receiver, the two hypotheses can be written as
\begin{equation}
\begin{array}{lll}
\hypzero: \obssig{t} &=& \recnoise{t},\\
\hypone: \obssig{t} &=& \propsig{t}+ \recnoise{t},\\
\end{array}
\label{specsenshypoth1}
\end{equation}
where $\recnoise{t}$ denotes receiver noise and $\propsig{t}$ stands for a \ac{pu} signal after propagation effects.

A number of spectrum sensing algorithms have been proposed in the literature \cite{yucek_survey_2009, zeng_review_2010, axell_spectrum_2012}.
There are three types of them, namely energy detection, stochastic feature detection and matched filter detection, where different types require different amounts of prior knowledge about the \ac{pu} signal.
While matched filter \cite[Ch. 4.3]{kay_fundamentals_1998} detectors require the knowledge of the exact waveform of at least part of the \ac{pu} signal\eg a pilot, energy detection \cite{urkowitz_energy_1967} does not require any prior knowledge. Feature detectors are an in-between as they only make assumptions about structural or statistical properties of the signal.

One of the stochastic features which lets an \ac{su} receiver discriminate between pure stationary noise (\hypzero) and a communication signal contaminated with noise (\hypone) is cyclostationarity.
In contrast to pure stationary noise, most man-made signals vary periodically with time \cite{gardner_exploitation_1991} and can thus be characterized as cyclostationary.
Although the data contained in a modulated signal may be a purely stationary random process, the coupling with sine wave carriers, pulse trains, repeating, spreading, hopping sequences and cyclic prefixes going along with its modulation causes a built-in periodicity \cite{gardner_signal_1988}.

One of the algorithms exploiting this fact for the purpose of spectrum sensing is the so called \emph{\ac{tdt}} as introduced in \cite{dandawate_statistical_1994}. The test can decide between the presence and absence of cyclostationarity for a pre-specified potential cycle frequency $\cycfreq$. It operates on the \ac{ca}, which, given an observed signal $\obssig{t}$, is defined as \cite{gardner_exploitation_1991}
\begin{equation}
\label{eq:symca}
\ca{}{}   = \liminfty{\obsdur} \frac{1}{\obsdur} \int\limits_{-\obsdur/2}^{\obsdur/2} \obssig{t + \delay/2} \obssigconj{t - \delay/2}  e^{-j2\pi\cycfreq t} \ud t
\end{equation}
for a potential cycle frequency $\cycfreq$ and a delay $\delay$.
For purely stationary signals $\ca{}{} = 0$ for all $\cycfreq \ne 0$, while for cyclostationary signals $\ca{}{} \ne 0$ for some $\cycfreq \ne 0$. The $\cycfreq$ with non-zero \ac{ca} coefficients are called cycle frequencies.
The set of cycle frequencies caused by one of potentially multiple incommensurate second-order periodicities in a cyclostationary signal comprises the periodicity's fundamental cycle frequency (the reciprocal of the fundamental period) as well as its harmonics (integer multiples).

Given the above information, we can rewrite the hypothesis test \ref{specsenshypoth1} as
\begin{equation}
\begin{array}{l}
\hypzero: \forall \{\cycfreq \in \realnum |\cycfreq \ne 0\} : \ca{}{} = 0,\\
\hypone: \exists \{\cycfreq \in \realnum |\cycfreq \ne 0\} : \ca{}{} \ne 0.\\
\end{array}
\label{eq:specsenshypoth2}
\end{equation}
Since the \ac{ca} is zero on its whole support except the set of cycle frequencies and $\cycfreq = 0$, it can be called sparse.
This sparsity can be taken advantage of for the purpose of estimating it from a small number of samples.

The exploitation of sparsity for signal recovery has a long history \cite{donoho_scanning_2010}.
The recent years, however, have seen a vastly accelerated development of the field resulting in a new sampling-paradigm called \ac{cs} \cite{donoho_compressed_2006,candes_robust_2006}.
Consider a time-series with a sparse discrete frequency spectrum\ie only few of the signal's frequency domain coefficients are non-zero. Clearly, the signal carries significantly less information than suggested by its size (number of coefficients). Given its support in the frequency domain it could be represented with far fewer coefficients.
However, looking at the time-series itself, it is not obvious that the signal could be compressed. Thus, in order to acquire the signal using traditional signal acquisition methods, a number of samples depending on the signal's \emph{dimension} rather than its \emph{information load} has to be taken. In contrast, applying \ac{cs} methods the signal can be recovered from a small subset of the otherwise required time-domain samples.

Multiple contributions have been made in the field of compressive cyclostationary spectrum sensing. 
The authors of \cite{khalaf_new_2014} formulate the estimation of the \acl{ca} as a sparse recovery problem, which they solve using the \ac{omp} algorithm \cite[Ch. 3.2]{foucart_mathematical_2013}.
Based on the sparse estimate of the \ac{ca}, they propose two detection algorithms exploiting different \ac{ca} properties. The first one, called \ac{scm}, compares the biggest \ac{ca} components \ac{omp} finds in two consecutive blocks of samples. If for both blocks the same discrete cycle frequencies are chosen, \hypone~is selected, otherwise \hypzero~is selected. The second detection algorithm is called \ac{sm}. It exploits the fact, that for certain types of signals, the \ac{ca} is symmetric around the \ac{dc} component. 
Instead of the \ac{ca}, the authors of \cite{tian_cyclic_2012} use the \ac{sc}, which is the Fourier transform of the \ac{ca} over $\delay$, for detecting multiple transmitters in a wideband signal using \acl{cs}. In order to estimate the \ac{sc} from compressed samples via \ac{cs}, they established a direct linear relation between the compressed samples and the \ac{sc}.
Based on \cite{tian_cyclic_2012}, the authors of \cite{rebeiz_cyclostationary-based_2012} derive a method for recovering the \ac{sc} from sub-Nyquist samples using a reduced complexity approach, for which they provide a closed form solution.
In \cite{cohen_cyclostationary_2011},  the \ac{mwc} \cite{mishali_theory_2010} is used to obtain the \ac{sc} from sub-Nyquist samples to then apply cyclostationarity detection.
Apart from cyclostationarity, \ac{cs} has seen utilization in different branches of spectrum sensing as for example in energy detection \cite{tian_compressed_2008,bollig_joint_2012,LaBoTh15}.

The contribution of this paper is manifold. 
We propose two novel \ac{ca} estimation algorithms, both of which exploit \emph{further} prior information about the \ac{ca} in addition to its sparsity: 
the simultaneous \ac{omp}-based (SOber) and the dictionary assisted (Dice) compressive \ac{ca} estimator. The first one exploits the joint sparsity of the \ac{ca} vectors with regard to the time delay in order to recover the \ac{ca} matrix for all delays simultaneously, while the second one takes advantage of the signal induced structure of the \ac{ca} by introducing structure dictionaries into the recovery process.
In order to evaluate the performance of the proposed \ac{ca} estimators we derive a closed-form expression of the \ac{ca} of \emph{sampled} linearly modulated signals with rectangular pulse shape.
Furthermore, we show how the expression can be used as prior information in the dictionary assisted approach.
Note, that the use of sparse recovery in the novel \ac{ca} estimation approaches results in the automatic detection of a signal's cycle frequencies. This in turn allows \emph{blind} spectrum sensing by eliminating the integral need of the classical \ac{tdt} for the perfect knowledge of said cycle frequencies. However, the resulting
sparse structure of the compressive \ac{ca} estimates does not allow for the application of the traditional \ac{tdt} since the noise statistics are missing.
To compensate for this phenomenon, we develop a modified \ac{tdt} and thus enable blind compressive cyclostationary spectrum sensing.
Numerical tests show that the proposed method achieves a near \ac{cfar} behavior.

The remainder of this paper is structured as follows.
\ref{sec:system_model} introduces the signal model and presents the classical method for \ac{ca} estimation.
In \ref{sec:tdt} the \acl{tdt} based on the classical \ac{ca} estimation is presented.
A \ac{ca} estimator based on joint sparsity of multiple vectors is introduced in \ref{sec:joint_omp}, while the \ac{ca} estimator exploiting additional prior knowledge is described in \ref{sec:cacafe}.
In \ref{sec:asyca} the closed-form \ac{ca} is derived.
An extension of the \ac{tdt} to accommodate sparse \ac{ca} estimates is developed in \ref{sec:sparse_tdt}.
The numerical evaluation of the proposed estimation and detection approaches as well as the interpretation of the results is given in \ref{sec:numerical_evaluation}.
\ref{sec:conclusion} concludes the paper.

\section{System Model and Classical \ac{ca} Estimation}
\label{sec:system_model}

Consider a secondary system receiver that needs to decide whether a certain spectral band is occupied or free. It samples the baseband signal $\obssig{t}$ uniformly with a sampling period $\smplper$.
This results in the vector of discrete samples $\dobssig{\timezero} \in \complexnum^\blocksize$, where
\begin{align}
\dobssig{\timezero} = [\obssig{\timezero}, \obssig{\timezero + \smplper}, \dotsc , \obssig{\timezero + (\blocksize-1)\smplper}]\T.
\end{align}
We assume the vector $\dobssig{\timezero}$ is discrete and zero-mean and due to the nature of man-made signals it represents an (almost \cite[Ch. 1.3]{napolitano_generalizations_2012}) cyclostationary process \cite{dandawate_statistical_1994}.
The presence of stochastic periodicity in the samples and thus the presence of a man-made signal can be revealed by applying a detection algorithm such as the \ac{tdt} to the \ac{ca} of the samples.
There are different ways of obtaining the \ac{ca} from the baseband samples, one of which is the following (classical) estimator
\begin{equation}
\label{eq:trad_est}
\caest{\dcycfreq}  = \frac{1}{\blocksize} \sum\limits_{n=0}^{\blocksize-1-\ddelay} \obssig{\timezero + n\smplper}
\obssigconj{\timezero + (n+ \ddelay) \smplper} e^{-j2\pi\frac{\dcycfreq}{\blocksize}n} e^{-j\pi\frac{\dcycfreq}{\blocksize}\ddelay}.
\end{equation}
Evaluating this function results in the \ac{ca} coefficient for the cycle frequency $\cycfreq = \frac{\dcycfreq}{\blocksize\smplper}$ and the time-delay $\delay = \ddelay \smplper$, where $\dcycfreq$ stands for the discrete cycle frequency and $\ddelay$ denotes the discrete time delay.
Note that the factor $e^{-j\pi\frac{\dcycfreq}{\blocksize}\ddelay}$ remains constant throughout the sum. It is a phase shift necessary to maintain compatibility with the symmetric \ac{ca} \ref{eq:symca}.
The estimator \ref{eq:trad_est} is biased but exhibits a smaller estimation variance than an unbiased one \cite{dandawate_statistical_1994}.

We define the \ac{ca} vector as 
\begin{align}
\cavec{} = [\caest{0}, \dotsc , \caest{\blocksize-1}]\T.
\end{align}
Subsequently, we rewrite the estimation of the \ac{ca} vector as a matrix-vector product.
To do so, we need the ($\blocksize$ element) \emph{delay-product} with time-delay $\delay = \ddelay \smplper$, which is given by 
\begin{align}
\label{delprod}
\delprodvec{\ddelay}{\blocksize} = \dobssig{\timezero} \circ \dobssigconj{\timezero + \ddelay \smplper},
\end{align}
where $\circ$ denotes component-wise multiplication.
Note that since the receiver only takes $\blocksize$ samples, $\dobssigconj{\timezero + \ddelay \smplper}$ is zero-padded at the end.
The \ac{ca} vector is now given by
\begin{align}
\cavec{\ddelay} = \frac{1}{\blocksize} \dftmat \delprodvec{\ddelay}{\blocksize},
\end{align}
where $\dftmat$ denotes the ($\blocksize \times \blocksize$) \ac{dft} matrix. 
The \ac{ca} matrix for time-delays $\ddelay_1\smplper, \dotsc , \ddelay_\ndelays\smplper$ is given by 
\begin{align}
\label{eq:trad_est_mat_vec}
\camat = [\cavec{\ddelay_1}, \dotsc , \cavec{\ddelay_\ndelays}] = \frac{1}{\blocksize} \dftmat \delprodmat{\blocksize},
\end{align}
with $\delprodmat{\blocksize} = [\delprodvec{\ddelay_1}{\blocksize}, \dotsc, \delprodvec{\ddelay_\ndelays}{\blocksize}]$.

\section{The \acf{tdt} for Cyclostationarity}
\label{sec:tdt}
As mentioned in \ref{sec:intro}, the statistical \ac{ca} of a cyclostationary signal is sparsely occupied, containing spikes only at the \ac{dc} component as well as the cycle frequencies of inherent signal periodicities and their harmonics.
Thus, given the statistical \ac{ca}, one could decide between \hypzero{} and \hypone{} by testing it for being non-zero at the signal's inherent  cycle frequencies. 
However, instead of the statistical \ac{ca}, we only have access to its estimation, the sample \ac{ca} (which asymptotically converges to the statistical \ac{ca}). 
The coefficients of the sample \ac{ca} are not constant but rather follow different probability distributions, depending on whether \hypzero{} or \hypone{} is true.
In the seminal work \cite{dandawate_statistical_1994}, these probability distributions have been identified and a test for cyclostationarity based on this knowledge has been designed. The test is briefly described in the following.

Consider the $1 \times 2\ndelays$ vector 
\begin{equation}
\begin{split}
\testvec(\atest) \hspace{-.2em}=\hspace{-.2em} \left[\hspace{-.1em}
\realpart{\camat[\atest,\ddelay_1]}, \hspace{-.1em}\dotsc\hspace{-.1em} , \realpart{\camat[\atest,\ddelay_\ndelays]}, \right. \\
\left. \imagpart{\camat[\atest,\ddelay_1]}, \hspace{-.1em}\dotsc\hspace{-.1em} , \imagpart{\camat[\atest,\ddelay_\ndelays]} 
\hspace{-.1em}\right],
\end{split}
\end{equation}
which represents the concatenation of the real and the imaginary part of the row of $\camat$ corresponding to the discrete cycle frequency $\atest$.
The frequency $\atest$ is the cycle frequency of interest\ie the one for the presence of which we want to test the signal.
Given this vector, we can formulate the following non-asymptotic hypotheses 
\begin{equation}
\begin{array}{lll}
\hypzero: \testvec(\atest) &=& \estnoise(\atest),\\
\hypone: \testvec(\atest) &=& \puretestvec(\atest) + \estnoise(\atest),\\
\end{array}
\label{specsenshypoth3}
\end{equation}
where $\puretestvec(\atest)$ is the deterministic but unknown asymptotic counterpart of $\testvec(\atest)$ and $\estnoise(\atest)$ is the estimation error.
Note that in contrast to the hypotheses from equation \ref{eq:specsenshypoth2}, this formulation considers the presence of cyclostationarity in the received signal for \emph{one fixed} cycle frequency $\atest$. 

Since $\puretestvec(\atest)$ is nonrandom, the distribution of $\testvec(\atest)$ under $\hypzero$ and $\hypone$ only differs in mean.
As shown in \cite{dandawate_statistical_1994}, the estimation error $\estnoise(\atest)$ asymptotically follows a Gaussian distribution\ie
\begin{equation}
\label{eq:est_error_zero_mean}
\liminfty{\blocksize} \sqrt{\blocksize} \estnoise(\atest) \convdist \normal{0}{\covmat(\atest)},
\end{equation}
where $\covmat(\atest)$ is the statistical covariance matrix of $\testvec(\atest)$ and $\convdist$ denotes convergence in distribution. 
The covariance matrix can be computed as \cite{dandawate_statistical_1994}
\begin{equation}
\covmat(\atest) = \left[ 
\begin{array}{cc}
\realpart{\frac{\qmat + \qmat^\ast}{2}} & \imagpart{\frac{\qmat - \qmat^\ast}{2}} \\
\imagpart{\frac{\qmat + \qmat^\ast}{2}} & \realpart{\frac{\qmat^\ast - \qmat}{2}}
\end{array}
\right],
\end{equation}
where the $(m,n)$-th entries of the matrices $\qmat$ and $\qmat^\ast$ are given by
\begin{align}
\qmat(m,n) &= \sc{2\atest}{\atest},~\text{and}\\
\qmat^\ast(m,n) &= \scconj{0}{-\atest}
\end{align}
respectively.
The term $\sc{\cdot}{\cdot}$ denotes the unconjugated, while the term $\scconj{\cdot}{\cdot}$ denotes the conjugated cyclic spectrum of a signal.
One way to estimate these is to determine the following frequency-smoothed periodograms:
\begin{equation}
\label{cycspec1}
\begin{array}{ll}
\estsc{2\atest}{\atest} = & \frac{1}{\blocksize L} \sum\limits_{s = - \frac{L-1}{2}}^{\frac{L-1}{2}}  W(s) \camat[\atest-s,\ddelay_n]\camat[\atest+s,\ddelay_m]
\end{array}
\end{equation}
\begin{equation}
\label{cycspec2}
\begin{array}{ll}
\estscconj{0}{-\atest} = & \frac{1}{\blocksize L} \sum\limits_{s = - \frac{L-1}{2}}^{\frac{L-1}{2}}  W(s) \camat^\ast[\atest+s,\ddelay_n]\camat[\atest+s,\ddelay_m],
\end{array}
\end{equation}
where $W$ is a normalized spectral window of odd length $L$.
Looking at the equations \ref{cycspec1} and \ref{cycspec2}, it becomes clear why the cyclic spectrum is often refered to as the spectral correlation.

Given the estimated quantities described above, the following generalized likelihood ratio (GLR) test statistic can be derived \cite{lunden_collaborative_2009}
\begin{equation}
\label{eq:dandaw_test_stat}
\teststat{\obssig{}\obssigconj{}} = \blocksize \testvec(\atest) \estcovmat^{-1}(\atest) \testvec\T(\atest).
\end{equation}
The test statistic can be interpreted as a normalized energy.
The inverse of the covariance matrix scales $\testvec(\atest)$ such that under $\hypzero$ its entries follow a standard normal distribution. Thus, under \hypzero{}, the test statistic asymptotically follows a central chi-squared distribution with $2\ndelays$ degrees of freedom\ie $\liminfty{\blocksize} \teststat{\obssig{}\obssigconj{}} \convdist \chi_{2\ndelays}^2$, while under \hypone{}, the test statistic asymptotically follows a non-central chi-squared distribution with unknown noncentrality parameter $\lambda$\ie $\liminfty{\blocksize} \teststat{\obssig{}\obssigconj{}} \convdist \chi_{2\ndelays}^2(\lambda)$.
Based on the above test statistic we can design a \ac{cfar} detector with some false alarm rate $\pfa$ by finding the corresponding decision threshold in the $\chi_{2\ndelays}^2$ tables.
We cannot design a test based on a desired detection rate $\pd$, since although $\puretestvec(\atest)$ is deterministic, it depends on the type of signal emitted by the transmitter as well as the \ac{snr} at the receiver, both of which are assumed to be unknown.

The classical approach for cyclostationary spectrum sensing is to apply the \ac{tdt} to the \ac{ca} estimate from \ref{eq:trad_est}. However, to do so one needs to know which cycle frequency to test beforehand, which eliminates the possibility of true \emph{blind} spectrum sensing. One could sequentially test the received signal for all possible cycle frequencies.
However, with high probability the estimation noise at some cycle frequency would have a value above the decision threshold, leading to a false alarm. 

\section{Simultaneous \ac{omp} \ac{ca} Estimation}
\label{sec:joint_omp}

In this section we cast the \ac{ca} estimation as a joint sparse recovery problem. Since this method is able to detect the \ac{ca}'s support, it removes the traditional approach's requirement of knowing the cycle frequencies beforehand, thus making truly blind cyclostationarity-based spectrum sensing possible.

We begin by rewriting equation \ref{eq:trad_est_mat_vec} as
\begin{equation}
\label{eq:inv_prob}
\delprodmat{\blocksize} = \blocksize \idftmat\camat,
\end{equation}
where $\idftmat$ is the \ac{idft} matrix.
Now consider an $\navailsmpls \times \blocksize$ matrix $\samplmat$, which consists of a selection of $\navailsmpls$ rows of the $\blocksize \times \blocksize$ identity matrix $\identitymat{\blocksize}$. 
It represents the undersampling operation. 
Applying $\samplmat$ to \ref{eq:inv_prob}, we get 
\begin{equation}
\label{eq:under_inv_prob}
\delprodmat{\navailsmpls} = \samplmat \delprodmat{\blocksize} = \blocksize \samplmat\idftmat\camat,
\end{equation}
where $\delprodmat{\navailsmpls}$ contains a selection of $\navailsmpls$ coefficients of the delay-products for different delays.
We now want to recover $\camat$ from $\delprodmat{\navailsmpls}$ by solving the underdetermined inverse problem \ref{eq:under_inv_prob}.
To do so we exploit our knowledge about the \ac{ca}'s sparsity.

The straightforward solution would be to solve the following optimization problem
\begin{equation}
\label{eq:lzeroproblem}
\begin{array}{rl}
\text{min}&\lpnorm{\vecop{\camat}}{0} \\
\text{s.t.}&\delprodmat{\navailsmpls} = \blocksize \samplmat\idftmat\camat,
\end{array}
\end{equation}
where $\lpnorm{\cdot}{0}$ denotes the ${\ell_0}$-``norm'' \cite{donoho_compressed_2006}, which is the number of non-zero entries in a vector, and $\text{vec}\{\cdot\}$ stands for the vectorization of a matrix\ie the concatenation of its columns to a single vector.
Equation \ref{eq:lzeroproblem} is known to be a non-convex combinatorial problem \cite{donoho_compressed_2006}. One way to solve it within a practically feasible amount of time is to substitute the ${\ell_0}$-``norm'' by its tightest convex relaxation, the $\ell_1$-norm. 
With high probability, this produces the same result since for most large underdetermined systems of linear equations the minimal $\ell_1$-norm solution is also the sparsest solution \cite{donoho_for_2004}.
Another way of solving \ref{eq:lzeroproblem} efficiently is applying one of the many greedy sparse recovery algorithms that have been developed in the field of \ac{cs}, such as\eg \emph{\acf{omp}}.

\ac{omp} is a greedy algorithm that iteratively determines a vector's support from an underdetermined system of linear equations and subsequently recovers the vector by solving a least-squares problem.
Using it, we could solve \ref{eq:lzeroproblem} for each column of $\camat$ individually (as in \cite{khalaf_new_2014})\ie we could solve
\begin{equation}
\begin{array}{rl}
\text{min}& \lpnorm{\cavec{\ddelay}}{0} \\
\text{s.t.}&\delprodvec{\ddelay}{\navailsmpls} = \blocksize \samplmat\idftmat\cavec{\ddelay},\\
\end{array}
\end{equation}
for each $\ddelay$. In order to exploit the additional knowledge that the vectors $\left.\cavec{\ddelay}\right|_{\ddelay = \ddelay_1}^{\ddelay_\ndelays}$ have the same support (they are jointly sparse with regard to the time delay), we propose to use an extension of \ac{omp} called \ac{somp} \cite{tropp_simultaneous_2005}.
The \ac{ca} estimation based on \ac{somp} is given in \ref{alg:somp}.
\begin{algorithm}[]
	\KwIn{$\delprodmat{\navailsmpls}, \niter, \atomsmat = \blocksize\samplmat\idftmat$}
	\KwOut{$\camat$}
	$\camat = {\bf 0}, S_0 = \emptyset$\;
	\For{$i = 1, \dotsc, \niter$}{
		$\corrmat = (\delprodmat{\navailsmpls} - \atomsmat\camat)\H \atomsmat$\;
		$S_i = S_{i-1} \cup \argmax{\lpnorm{\corrvec{j}}{1}}{j \in \{1, \dotsc, \blocksize\}}$\;
		\For{$k = 1, \dotsc, \ndelays$}{
			$\cavec{\ddelay_k} = \argmin{\lpnorm{\delprodvec{\ddelay_k}{\navailsmpls} - \atomsmat{\bf z}}{2}, \supp{{\bf z}} \subset S_i}{{\bf z} \in \complexnum^\blocksize}$\;
		}
	}
	\caption{\ac{somp}\cite{tropp_simultaneous_2005}-based \ac{ca} estimator (\sober)}
	\label{alg:somp}
\end{algorithm}
The number of iterations is denoted by $\niter$, the $\ndelays \times \blocksize$ matrix $\corrmat$ contains correlation values and $\corrvec{j}$ is its $j$th column, $\delprodvec{\ddelay_k}{\navailsmpls}$ denotes the $k$th column of $\delprodmat{\navailsmpls}$ and $\supp{\cdot}$ stands for the support of a vector\ie the indices of its non-zero entries.
We refer to the columns of the matrix $\atomsmat = \blocksize\samplmat\idftmat$ introduced in \ref{alg:somp} as atoms. 

Since asymptotically, $\camat$ has only few rows with non-zero entries, up to a certain residuum the columns of $\delprodmat{\navailsmpls}$ should be representable by a weighted combination of only few of the atoms contained in $\atomsmat$. The goal of the algorithm is to find the indices of the atoms contained in $\delprodmat{\navailsmpls}$\ie the support of the columns of $\camat$, and subsequently recover the identified non-zero rows of $\camat$ by solving least-squares problems. We start with an empty support $S_0$. 
Each iteration, one atom index is added to the support.
The index is selected according to the sum of the absolute correlation values between the corresponding atoms and the delay products of different time delays (lines 3-4).
Using the new support set $S_i$, a least-squares problem is solved for each column in $\camat$ (lines 5-6).
In each iteration the atom index to be added to the index set is  chosen according to the correlation between the residuum of $\delprodmat{\navailsmpls}$ and the atom set.
Since every iteration adds one index to the support set, one usually chooses $\niter$ greater than or equal to the sparsity of the signal to be recovered.
The difference between \ac{omp} (used in\eg \cite{khalaf_new_2014}) and \ac{somp} can be found in line 4, where \ac{somp} jointly considers the amount of correlation between atoms and the delay products of \emph{multiple} delays, while \ac{omp} would select the support of $\left.\cavec{\ddelay_l}\right|_{l=1}^\ndelays$ for each $l$ individually. 

The support determined by the algorithm constitutes a set of cycle frequencies. To determine the observed band's occupancy status, we want to test for the presence of cyclostationarity at the support using the \ac{tdt}.
However, since only few of the coefficients of $\camat$ are recovered and all other coefficients are set to zero, it is not possible to estimate the covariance matrix $\estcovmat$ as part of the \ac{tdt} as presented in \ref{sec:tdt}. 
To tackle this problem, a modified \ac{tdt} is presented in \ref{sec:sparse_tdt}.

\section{Dictionary Assisted \ac{ca} Estimation}
\label{sec:cacafe}

In \ref{sec:joint_omp} we have described a \ac{somp}-based algorithm that estimates the cycle frequencies and the \ac{ca} from fewer samples than required using the classic approach by taking into consideration the inherent sparsity of the \ac{ca}.
In this section we develop an algorithm that makes use of additional prior knowledge about the signal's structure in the form of structure dictionaries to further enhance the cycle frequency and \ac{ca} estimation.
Like \sober, the new algorithm does not require the prior knowledge about the cycle frequencies contained in the signal.

One fact about the \ac{ca} that could be exploited 
is that using a rectangular pulse shape, a linearly modulated signal's \ac{ca} exhibits spikes not only at the signal's fundamental cycle frequency but also at the harmonics thereof.
Another one is the symmetry of the \ac{ca} around the DC component. 
First steps in this direction showing promising results have been taken in \cite{bollig_dictionary-based_2013}. The drawback of the solution proposed in \cite{bollig_dictionary-based_2013} is that the convex optimization problem used to recover the \ac{ca} becomes huge for practical parameter choices, which results in a prohibitively large computational complexity. To circumvent this we propose an \ac{omp}-based greedy algorithm that takes advantage of the additional prior knowledge while featuring a much smaller complexity than the optimization problem.
In the following we introduce a structure dictionary accounting for the symmetry of the \ac{ca} and describe the proposed dictionary assisted recovery algorithm. In the next section we discuss a second structure dictionary that can be used with the proposed algorithm, i.e., the dictionary containing the harmonic structure of the \ac{ca} as well as its shape.

Let $\symdict{\frac{\blocksize}{2}} \in \{0,1\}^{\frac{\blocksize}{2}\times\frac{\blocksize}{2}}$ denote the symmetry dictionary. Its columns represent possible cycle frequencies contained in the set $\dcycfreq \in \{1, \dotsc, \frac{\blocksize}{2}\}$.
For simplicity, this set is chosen such that the frequencies contained in it lie at the center frequencies of the \ac{ca}'s \ac{dft} bins.
An entry of the dictionary covers elements $1$ to $\frac{\blocksize}{2}$ of $\cavec{\ddelay}$ which is indexed from $0$ to $\blocksize-1$.
The symmetry dictionary is simply given by the identity matrix\ie $\symdict{\frac{\blocksize}{2}} = \identitymat{\frac{\blocksize}{2}}$.
To model the whole vector $\cavec{\ddelay}$, the dictionary is extended to include the \ac{dc} component, which is set to zero, as well as the negative cycle frequencies.
Note that the \ac{dc} component is set to zero 
because its value is independent of the presence of cyclostationarity.
The resulting \emph{full} dictionary is exemplarily given by
\begin{equation}
\symdictf{3} = \left(\begin{array}{ccc}
0   & 0   & 0\\
1   & 0   & 0\\
0   & 1   & 0\\
0   & 0   & 1\\
0   & 1   & 0\\
1   & 0   & 0\\
\end{array}\right).
\end{equation}
The circle above the symbol indicates that it is the full version of the dictionary\ie the one spanning the whole Fourier range.
The \emph{ones} in the matrix specify the locations of the non-zero coefficients in the \ac{ca} fitting the format of \ref{eq:trad_est_mat_vec}.

\begin{algorithm}[]
	
	\KwIn{$\delprodmat{\navailsmpls}, \niter, \atomsmat = \blocksize\samplmat\idftmat, \structdictf{\frac{\blocksize}{2}}_l|_{l=1}^\ndelays$}
	
	\KwOut{$\camat$}
	
	$\camat = {\bf 0}, S_0 = \{0\}$\; 
	
	\For{$i = 1, \dotsc, \niter$}{
		
		\For{$l = 1, \dotsc, \ndelays$}{
			$\left[\ringcorrmat\right]_{l:} = \operatorname{abs}\left( (\delprodvec{\ddelay_l}{\navailsmpls} - \atomsmat \cavec{\ddelay_l})\H \atomsmat \right) \structdictf{\frac{\blocksize}{2}}_l$\;
		} 

		$S_i = S_{i-1} \cup \left\{h \left| [\structdictf{\frac{\blocksize}{2}}]_{hj} \neq 0, j = \argmax{
			\lpnorm{\ringcorrvec{j}}{1}}{j \in \{1, \dotsc, \frac{\blocksize}{2}\}   
		} \right. \right\}$\;
		
		\For{$k = 1, \dotsc, \ndelays$}{
			$\cavec{\ddelay_k} = \argmin{ \lpnorm{ \delprodvec{\ddelay_k}{\navailsmpls} - \atomsmat{\bf z} }{2}, \supp{\bf z} \subset S_i }{ {\bf z} \in \complexnum^\blocksize }$\;
		}
	}
	
	\caption{Dictionary assisted \ac{ca} estimator (\hades)}
	\label{alg:cacafe}
\end{algorithm}

The \hades{} algorithm (\ref{alg:cacafe}) follows the same idea as the \sober{} algorithm (\ref{alg:somp}) in that it iteratively determines the support of the sparse \ac{ca} and subsequently recovers it by solving an overdetermined least-squares problem. 
However, in contrast to \sober{}, \hades{} facilitates the use of further prior knowledge in addition to the \ac{ca}'s sparsity in the recovery process.

Thus, in addition to the inputs received by \sober{}, \hades{} needs a set of structure dictionaries $\structdictf{\frac{\blocksize}{2}}_l|_{l=1}^\ndelays$, one dictionary for each delay value $\ddelay_l$ that is to be considered in the recovery process.
In the case of the symmetry dictionary, all of these are identical\ie $\structdictf{\frac{\blocksize}{2}}_l|_{l=1}^\ndelays = \symdictf{\frac{\blocksize}{2}}$.
Since the structure dictionaries do not necessarily model the DC component of the \ac{ca}, it is added to the support set in the initialization phase in \hades{} (line 1).
Instead of working with the amount of correlation between the residuum and the atoms directly as in \sober{}, the \hades{} algorithm computes combinations of these as dictated by the structure dictionaries in use (lines 3, 4). 
This way, the decision about the non-zero cycle frequencies (line 5) takes into account the structure of the \ac{ca}.
Additionaly, instead of adding a single element to the support set per iteration, \ref{alg:cacafe} adds all indices to the support set that have a non-zero value in the selected dictionary word.
The recovery step (cf. lines 6, 7) remains unchanged.
Note that in \ref{alg:cacafe} the $\operatorname{abs}(\cdot)$ operator stands for the element-wise absolute value of a matrix, while the selection operator $[\cdot]_{l:}$ denotes the $l$th row of a matrix.

\section{Asymptotic \ac{ca} and asymptotic dictionary}
\label{sec:asyca}

The symmetry structure dictionary exploits one of the facts we know about the \ac{ca}. In order to explore an extreme in terms of prior knowledge 
we create a dictionary that contains the maximum possible amount of prior information about the \ac{ca}\ie the one containing the asymptotic \ac{ca} itself. 
This requires knowledge of the analytic expression for the discrete asymptotic \ac{ca} vector, which we derive in the following.

To assess the performance of different \ac{ca} estimation algorithms we employ common linearly modulated signals with symbol length $\symblen$ as described by the following equation \cite[Eq. 73]{gardner_exploitation_1991}
\begin{equation}
\label{eq:transsig}
\transsig{t} = \sum\limits_{n = -\infty}^\infty \symb_n \pulseshape{t - n\symblen + \phi}.
\end{equation}
Here, $\pulseshape{t}$ is a deterministic finite-energy pulse, $\phi$ represents a fixed pulse-timing phase parameter and $\symb_n$ stands for the $n$-th symbol to be transmitted.
We are now interested in an expression for the discrete asymptotic \ac{ca} vector of the above signal type.

The fundamental cycle frequency of the built-in periodicity of the signal from \ref{eq:transsig} is $\frac{1}{\symblen}$.
Its continuous \ac{ca} is given by \cite[Eq. 81]{gardner_exploitation_1991}
\begin{equation}
\label{eq:caofs}
\transca{}{} = \left\{ \begin{array}{ll}
0 & \text{for}~\cycfreq \ne \frac{k}{\symblen}\\
\frac{1}{\symblen} \sum\limits_{n = -\infty}^\infty \randsigca{n \symblen}{} \pulseshapeca{\delay - n\symblen} e^{j2\pi \cycfreq \phi} & \text{otherwise,}
\end{array}\right.
\end{equation}
where $k \in \integers$ and $\pulseshapeca{\delay}$ is defined as \cite[Eq. 82]{gardner_exploitation_1991}
\begin{equation}
\label{eq:psca}
\pulseshapeca{\delay} \triangleq \int\limits_{-\infty}^\infty \pulseshape{t + \delay / 2}\conjpulseshape{t - \delay / 2} e^{-j2\pi \cycfreq t} \ud t.
\end{equation}
The symbol $\integers$ denotes the set of integers\ie $k \in \{\dotsc, \shortminus2, \shortminus1, 0, 1, 2, \dotsc\}$.

We consider the case where $\symb_n$ is a purely stationary random sequence. Thus, its autocorrelation $\randsigca{n \symblen}{} = \randsigca{n \symblen}{0}$ is non-zero only at $n = 0$ (cf. \ref{eq:symca}), reducing \ref{eq:caofs} to
\begin{equation}
\label{eq:caofssimpl}
\transca{}{} = \left\{ \begin{array}{ll}

0 & \text{for}~\cycfreq \symblen \notin \integers\\
\frac{\sigpow{}}{\symblen} \pulseshapeca{\delay} e^{j2\pi \cycfreq \phi} & \text{otherwise,}

\end{array}\right.
\end{equation}
where $\sigpow{}$ is the average power of $\symb_n$.
In the following we assume a rectangular pulse shape of length $\symblen$\ie $\pulseshape{t} = \operatorname{rect}(\frac{t}{\symblen})$, which leads to $\pulseshape{t + \frac{\delay}{2}}\conjpulseshape{t - \frac{\delay}{2}} = \operatorname{rect}\left(\frac{t}{\symblen - |\delay|}\right)$. Thus, applying the Fourier transform to \ref{eq:psca} yields
\begin{equation}
\label{eq:caofssimplrect}
\transca{}{} = \left\{ \begin{array}{ll}

0 & \text{for}~\cycfreq \symblen \notin \integers\\
\sigpow{} \frac{\symblen - |\delay|}{\symblen} \operatorname{sinc}(\cycfreq (\symblen - |\delay|)) e^{j2\pi \cycfreq \phi} & \text{otherwise,}

\end{array}\right.
\end{equation}
%
%
for $|\delay| \le \symblen$ where 
$\operatorname{sinc}(x) = \frac{\operatorname{sin}(\pi x)}{\pi x}$.
Note that the use of the absolute value of the delay stems from the fact that for a real symmetric pulse shape $\pulseshape{t}$, the expression $\pulseshape{t + \frac{\delay}{2}}\conjpulseshape{t - \frac{\delay}{2}}$ is symmetric with respect to $\delay$.

Equation \ref{eq:caofssimplrect} represents the \ac{ca} of the continuous-time signal described by \ref{eq:transsig}. The \ac{ca} of the sampled version of \ref{eq:transsig} at its fundamental cycle frequency and the harmonics thereof is given by \begin{equation}
\label{eq:finalanalca}
\left.\dca{\dcycfreq}{\ddelay}{\nsym}\right|_{\dcycfreq = k\frac{\blocksize}{\nsym}} = 
\frac{\sigpow{}}{\nsym} \frac{\operatorname{sin}(\pi \frac{\dcycfreq}{\blocksize}(\nsym - |\ddelay|))}{\operatorname{sin}\left( \pi \frac{\dcycfreq}{\blocksize} \right)} e^{j2\pi \frac{\dcycfreq}{\blocksize}d_\phi}.
\end{equation}
The derivation of this expression can be found in the appendix.

The coefficients of the closed-form expression \ref{eq:finalanalca} together with the alternative case $\left.\dca{\dcycfreq}{\ddelay}{\nsym}\right|_{\dcycfreq \ne k\frac{\blocksize}{\nsym}} = 0$ at different discrete cycle frequencies $\dcycfreq$ are arranged in a vector $\analcavec{\nsym}{\dcycfreq}{\ddelay}$ matching the format of the \ac{dft} matrix, such that
\begin{equation}
\label{eq:th_vec}
\analcavec{\nsym}{\dcycfreq}{\ddelay} \hspace{-.2em}=\hspace{-.2em} \left\{ \hspace{-.7em}\begin{array}{ll}
{ \dca{\dcycfreq}{\ddelay}{\nsym} }&\hspace{-.63em}\text{for}~\dcycfreq \in \{0, \dotsc ,\frac{\blocksize}{2}\},\\
{ \dca{(\dcycfreq - \blocksize)}{\ddelay}{\nsym} }&\hspace{-.63em}\text{for}~\dcycfreq \in \{\frac{\blocksize}{2}+1, \dotsc ,\blocksize-1\}.
\end{array}\right.
\end{equation}
Note, that adding purely stationary noise to the signal $\transsig{t}$ does not change its asymptotic \ac{ca} (with the exception of $(\dcycfreq, \ddelay) = (0,0)$, at which point the \ac{ca}'s value is the average power of signal and noise, cf. \ref{eq:symca}) since the noise exhibits no inherent periodic behaviour. 
Due to this fact, \ref{eq:th_vec} can also be used as a reference for the \ac{ca} of signals contaminated with \ac{awgn} with the exception mentioned.

Given \ref{eq:th_vec} we can now construct the asymptotic dictionary:
\begin{equation}
\asydictf{\frac{\blocksize}{2}} = \left[
\frac{\operatorname{abs}\left( \analcavec{ \nsym = \frac{\blocksize}{1} }{}{\ddelay_l} \right)}{\lpnorm{\analcavec{\nsym = \frac{\blocksize}{1} }{}{\ddelay_l}}{1}}
,\dotsc,
\frac{\operatorname{abs}\left( \analcavec{ \nsym = \frac{\blocksize}{\nicefrac{\blocksize}{2}} }{}{\ddelay_l} \right)}{\lpnorm{\analcavec{\nsym = \frac{\blocksize}{\nicefrac{\blocksize}{2}} }{}{\ddelay_l}}{1}}
\right].
\end{equation}
Note, that in contrast to the single symmetry dictionary, there is a whole set of asymptotic dictionaries, one for each delay value of interest. The columns of the dictionaries correspond to actual symbol lengths\ie actual cycle frequencies. Thus, each column contains the absolute value of the normalized asymptotic \ac{ca} of a cycle frequency candidate where the discrete symbol lengths $\nsym \in \left\{ \frac{\blocksize}{1}, \dotsc, \frac{\blocksize}{\nicefrac{\blocksize}{2}} \right\}$ correspond to the discrete cycle frequencies $\dcycfreq \in \left\{ 1, \dotsc, \nicefrac{\blocksize}{2} \right\}$.
It is worth noting that in addition to its role as the basis of the second structure dictionary for \ref{alg:cacafe}, the expression \ref{eq:th_vec} serves as a reference for the direct comparison of different \ac{ca} estimation methods in \ref{sec:numerical_evaluation}.

\section{Cyclostationarity Detection from Sparse Cyclic Spectra}
\label{sec:sparse_tdt}
Both, the \ac{somp}-based (\ref{alg:somp}) and the dictionary assisted \ac{ca} estimation (\ref{alg:cacafe}), are able to recover the \ac{ca} without knowing which cycle frequencies are contained in the signal beforehand. 
However, although this makes for a good \ac{ca} estimation, it is not directly compatible with the traditional \ac{tdt} described in \ref{sec:tdt}, since it only recovers the \ac{ca} coefficients at the cycle frequencies.

The \ac{tdt} is a \ac{cfar} detector\ie the \ac{pdf} of its test statistic under $\hypzero$  is asymptotically independent of any signal parameters like\eg the noise power. 
To achieve this, the \ac{tdt} first estimates the \ac{ca} noise covariance and then rescales the original \ac{ca} by this estimate so that the scaled \ac{ca} follows a \emph{standard} Gaussian distribution. 
This is where the problem occurs. Although, we are ultimately only interested in the \ac{ca} coefficients that are located at the signal's cycle frequencies, for the estimation of the noise covariance we need the 
coefficients lying between the cycle frequencies, which only carry estimation noise. 
\sober{} and \hades{} do not recover these. Thus, we propose an extension to the \ac{tdt}, the sparse \ac{tdt}, to bridge this gap in the following.

To obtain optimal \ac{ca} recovery performance one would choose the sensing matrix $\atomsmat$ with minimum structure\ie the selection of the $\navailsmpls$ entries of the delay product would be completely random.
However, to tackle the aforementioned problem we choose a combination of consecutive and random delay product elements. 
The consecutive part comprises the first $\lceil\consecfrac\navailsmpls\rceil$ rows of $\delprodmat{\navailsmpls}$, where $\consecfrac \in [0.01,0.5]$ and $\lceil\cdot\rceil$ denotes the ceiling operation. The remainder of the rows of $\delprodmat{\navailsmpls}$ is a random selection of the remaining rows of $\delprodmat{\blocksize}$.
The first step of the sparse \ac{tdt} is to determine the classical \ac{ca} estimation of the consecutive block of delay product elements.
In the next step the cycle frequency of interest $\atest$ is determined using either \ref{alg:somp} or \ref{alg:cacafe}.
Next, the covariance matrix for the cycle frequency $\atest$ corresponding to the $\blocksize$-size \ac{ca} $\left(\estcovmat^{(\blocksize)}(\atest)\right)$ needs to be determined, where the superscript $(\blocksize)$ indicates the corresponding \ac{ca} size.
It is obtained as
\begin{equation}
\estcovmat^{(\blocksize)}(\atest) = \frac{\estcovmat^{(\lceil\consecfrac\navailsmpls\rceil)}(\lceil\consecfrac\frac{\navailsmpls}{\blocksize}\atest\rceil)}{\sqrt{\consecfrac\frac{\navailsmpls}{\blocksize}}},
\end{equation}
where $\estcovmat^{(\lceil\consecfrac\navailsmpls\rceil)}$ is the covariance matrix corresponding to the $\lceil\consecfrac\navailsmpls\rceil$-size \ac{ca} estimated from the consecutive samples in the first step.
The test statistic is subsequently evaluated as (cf. \ref{eq:dandaw_test_stat})
\begin{equation}
\teststat{\obssig{}\obssigconj{}}^\text{\,sparse} = \blocksize \testvec(\atest) \left( \frac{\estcovmat^{(\lceil\consecfrac\navailsmpls\rceil)}(\lceil\consecfrac\frac{\navailsmpls}{\blocksize}\atest\rceil)}{\sqrt{\consecfrac\frac{\navailsmpls}{\blocksize}}} \right)^{-1}\hspace{-.5em} \testvec\T(\atest).
\label{eq:testStat}
\end{equation}

The consecutive sample ratio $\consecfrac$ is a trade-off parameter. The optimal sparse recovery performance is to be expected for the case that $\atomsmat = \blocksize\samplmat\idftmat$ has the smallest possible amount of structure, which here corresponds to the case where the set of known delay product elements is chosen completely at random\ie for $\consecfrac = 0$. Contrarily, the best estimation quality for the \ac{ca} covariance matrix $\estcovmat$ is achieved when all known delay product elements are consecutive\ie for $\consecfrac = 1$. 

\section{Numerical Evaluation}
\label{sec:numerical_evaluation}

In this section we compare the performance of the methods presented in the preceding sections. The parameters used throughout this section are given in \ref{tab:sim_par}.

We begin by investigating the influence of the consecutive sample ratio $\consecfrac$ on the spectrum sensing performance. \ref{fig:pd_over_consec} shows how the detection rate changes with $\consecfrac$ for an \ac{snr} of $0$ dB and different false alarm rates.
For all methods but the \ac{omp}, $\consecfrac = 0.15$ seems to be a good choice. For the \ac{omp} the detection rate increases monotonically with $\consecfrac$. However, as can be seen below, even for the \ac{omp}, a high $\consecfrac$ is no good choice regarding other performance categories.

In \ref{fig:max_pd_over_snr} the best achievable detection rate\ie the detection rate for the individual best choice of $\consecfrac$, of the different detectors is plotted over the receiver \ac{snr} for different false alarm rates. The term \emph{oracle} expresses that a method has prior knowledge about the exact cycle frequencies contained in the signal. The classic method depends on this knowledge while for the sparse recovery, it reduces the \ac{ca} recovery to solving the overdetermined least squares problem for the given support (cf. lines 5 and 6 in \ref{alg:somp} or lines 6 and 7 in \ref{alg:cacafe}).
As expected, the oracle methods outperform the methods which have to determine the \ac{ca} support themselves by a large margin. Regarding the case of missing support knowledge, the \hades{} algorithm clearly outperforms the \sober{} algorithm as well as \ac{omp}.
It is to be noted that both, \ref{fig:pd_over_consec} as well as \ref{fig:max_pd_over_snr} do not show a significant performance advantage of exploiting the full knowledge of the asymptotic \ac{ca} (\hades{} (asy)) over just exploiting its symmetry property (\hades{} (sym)) for a sensible choice of $\consecfrac$.

The lines in \ref{fig:far_settings} show which false alarm rate according to the ideal chi-squared distribution has to be set in order to achieve 1, 3, 5, and 10 percent false alarm rate in the actual system. The dashed lines cross at the desired false alarm rate with $\consecfrac = 0.15$. While the two \hades{} methods roughly keep within a one percent offset, \ac{omp} and \sober{} show a decreasing degree of equivalence for an increasing false alarm rate. 
This indicates that using the chi-squared distribution for setting the decision threshold of the \hades{} algorithm is viable, which is an important observation. It means that in contrast to many other spectrum sensing algorithms, \hades{} approximately possesses a desirable feature called constant false alarm rate\ie its test statistic is independent of system parameters like the receiver noise power.

\ref{fig:hitrates_absidxerrors} shows how well the support of the \ac{ca} is recovered by the different methods. Since different types of communication signals feature different cycle frequencies, this information can be used for system identification. The hitrate is the chance of exactly recovering the correct support while the absolute index error is the mean recovery error in terms of \ac{ca} bins. Obviously, the \hades{} methods have superior support recovery capabilities.

The final performance category we evaluate is the \ac{ca} estimation quality achievable by sparse recovery methods measured by the \ac{mse}. In the left graph of \ref{fig:caf_mse} the \ac{mse} over the whole \ac{ca} is plotted while the right graph shows the \ac{mse} at the spikes of the \ac{ca}\ie the \ac{mse} at the actual cycle frequencies. To determine the error, we use the analytic expression for the asymptotic \ac{ca} vector as derived in \ref{sec:asyca}\ie the \ac{mse} is defined as
\begin{equation}
\frac{\|\camat - \left[ \analcavec{\nsym}{}{\ddelay_1}, \dotsc, \analcavec{\nsym}{}{\ddelay_\ndelays} \right]\|_F^2}{\blocksize\ndelays},
\end{equation}
where $\|\cdot\|_F$ is the Frobenius norm.
The sparse recovery method has a much lower \emph{overall} \ac{mse}. This is caused by the fact that it sets all \ac{ca} coefficients but the detected support to zero while the classical method results in a \ac{ca} that features estimation noise between the spikes.
Regarding the \emph{spike} \ac{mse}, both methods seem to perform roughly equivalently.

\section{Conclusion}
\label{sec:conclusion}
Blind operation and constant false alarm rate (CFAR) are desirable characteristics of spectrum sensing algorithms. Unfortunately, cyclostationarity-based approaches typically only feature either one or the other. We showed that this can be changed by using sparse recovery methods in the \ac{ca} estimation.
Subsequently, we developed a way to use further prior knowledge in addition to sparsity for superior \ac{ca} estimation. We derived a closed-form expression of the \ac{ca} of \emph{sampled} linearly modulated signals with rectangular pulse shape to be used both as prior information for the \ac{ca} estimation and as a reference for comparison. Finally, we extended a well known statistical test for cyclostationarity to accommodate sparse input. The results allow us to conclude that the proposed \hades{} algorithm in combination with the symmetric structure dictionary constitutes a viable alternative to the classical \ac{tdt} for the case of missing prior information about the cycle frequencies contained in the signal.

\section*{Appendix: Discrete Asymptotic \ac{ca}}
\label{app:dasyca}

The relation between the \ac{ca} of the continuous time-domain signal $\transsig{t}$ and its sampled counterpart $\{\transsig{n\smplper}\}$ is given by \cite[Ch. 11, Sec. C, Eq. 111]{gardner_statistical_1986}
\begin{equation}
\label{eq:timesmpl}
\begin{array}{l}
\dca{}{}{} = \sum\limits_{l=- \infty}^{\infty} \transca{\cycfreq + \frac{l}{\smplper}}{\ddelay\smplper} e^{j\pi l \ddelay}.
\end{array}
\end{equation}
The sum over $l$ reflects the infinite aliasing caused by the sampling.
In the next step, we insert \ref{eq:caofssimplrect} into \ref{eq:timesmpl}. 
Also we express quantities in terms of the sampling period $\smplper$\ie $\symblen \rightarrow \nsym \smplper, \cycfreq \rightarrow \frac{\dcycfreq}{ \blocksize \smplper}, \phi \rightarrow d_\phi \smplper$, 
with $\nsym, \dcycfreq, \blocksize \in \integers$.
This leads to
\begin{equation}
\label{eq:fullydiscreteca}
\dca{\dcycfreq}{\ddelay}{\nsym} \hspace{-.2em}=\hspace{-.2em} \left\{ \hspace{-.4em}\begin{array}{ll}
0 & \hspace{-.6em}\text{for}~\dcycfreq\!\ne\!\frac{k\blocksize}{\nsym}\\
\sigpow{} \frac{\nsym - |\ddelay|}{\nsym} e^{j2\pi \frac{\dcycfreq}{\blocksize}d_\phi}   \sum\limits_{l=- \infty}^{\infty} e^{j\pi l\ddelay} &\\
\cdot e^{j2\pi l d_\phi} \operatorname{sinc}((\frac{\dcycfreq}{\blocksize} + l) (\nsym - |\ddelay|)) &\hspace{-.6em}\text{otherwise,}
\end{array}\hspace{-2em}\right.
\end{equation}
for $|\ddelay| \le \nsym$, where $\nsym$ is the oversampling factor.
In this step we used the fact that for our assumptions all aliases of the fundamental cycle frequency and its harmonics lie on top of the actual fundamental cycle frequency and its harmonics\ie $(\cycfreq + \frac{l}{\smplper}) \symblen \in \integers$ iff $\cycfreq\symblen \in \integers$.
Inserting the discrete quantities given above, we get $(\frac{\dcycfreq}{\blocksize} + l) \nsym \in \integers$ iff $\frac{\dcycfreq}{\blocksize} \nsym \in \integers$.
Since $\nsym \in \integers$ and $l \in \integers$, this always holds.
To rule out any spectral leakage, we choose $\blocksize$ as an integer multiple of $\nsym$, since then $\dcycfreq = k \frac{\blocksize}{\nsym}$ is also an integer and thus the fundamental discrete cycle frequency and its harmonics hit center frequencies of frequency bins. 
%
%
%
%

For $\dcycfreq = k\frac{\blocksize}{\nsym}$ expression \ref{eq:fullydiscreteca} can be shown to be
\begin{equation}
\label{eq:fullydiscreteca2}
\hspace{-.2em}\begin{array}{l}
\left.\dca{\dcycfreq}{\ddelay}{\nsym}\right|_{\dcycfreq = k\frac{\blocksize}{\nsym}} \hspace{-.2em}=\hspace{-.2em} 
\sigpow{} \frac{\operatorname{sin}(\pi \frac{\dcycfreq}{\blocksize}(\nsym - |\ddelay|))}{\pi\nsym} e^{j2\pi \frac{\dcycfreq}{\blocksize}d_\phi}  \hspace{-.8em}\sum\limits_{l=- \infty}^{\infty} \hspace{-.4em}\frac{(-1)^{l}}{\frac{\dcycfreq}{\blocksize} + l}.
\end{array}\hspace{-1em}
\end{equation}
To obtain \ref{eq:fullydiscreteca2} we used the definition of the $\operatorname{sinc}$ and exploited the facts that $e^{j\pi k} = (-1)^{k}$ for $k \in \integers$ and that $\operatorname{sin}(x + k\pi) = (-1)^{k}\operatorname{sin}(x)$ for $k \in \integers$.
The pulse timing phase parameter $d_\phi$ was set to $\frac{\nsym + 1}{2}$. This has the following reason. In order to simplify the numerical evaluation, we want to choose $\phi$ such that the beginning of the observed receiver signal is aligned with the rectangular pulse shapes\ie we would set $\phi = \frac{\symblen}{2}$. However, doing so would lead to the need to sample at the discontinuities caused by the instant change in amplitudes at the transition between symbols. To avoid this we choose $\phi = \frac{\symblen}{2} + \epsilon$, where $\epsilon \in (0, \smplper)$.
Note, that \ref{eq:fullydiscreteca2} is the same for any $\epsilon \in (0, \smplper)$.
In order to ease the derivation we can thus choose $\epsilon = \frac{\smplper}{2}$\ie $d_\phi = \frac{\nsym + 1}{2}$.

The infinite series in \ref{eq:fullydiscreteca2} can be expressed as
\begin{equation}
\label{eq:unevenone}
\hspace{-.2em}{
	\begin{array}{l}
	\sum\limits_{l=- \infty}^{\infty} \frac{(-1)^{l}}{\frac{\dcycfreq}{\blocksize} + l} 
	= \frac{\blocksize}{\dcycfreq} + \sum\limits_{l=1}^{\infty}
	\frac{(-1)^{l}}{\frac{\dcycfreq}{\blocksize} + l}
	+ \frac{(-1)^{l}}{\frac{\dcycfreq}{\blocksize} - l}
	\\
	= \frac{\blocksize}{\dcycfreq} + \frac{1}{2} \sum\limits_{l=1}^{\infty}
	-\frac{1}{l + \anhalf - \half} 
	+ \frac{1}{l - \anhalf 
		- \half} - \frac{1}{l - \anhalf} 
	+ \frac{1}{l + \anhalf}\\
	= \frac{\blocksize}{\dcycfreq} + \frac{1}{2} \sum\limits_{l=1}^{\infty}
	- \frac{l + \anhalf - \half - \anhalf + \half}{l(l + \anhalf - \half)}
	+ \frac{l - \anhalf - \half + \anhalf + \half}{l(l - \anhalf - \half)}\\
	\hspace{4.5cm}- \frac{l - \anhalf + \anhalf}{l(l - \anhalf)}
	+ \frac{l + \anhalf - \anhalf}{l(l + \anhalf)} \\
	= \frac{\blocksize}{\dcycfreq} + \frac{1}{2} \sum\limits_{l=1}^{\infty}
	\frac{\anhalf - \half}{l(l + \anhalf - \half)}
	- \frac{- \anhalf - \half}{l(l - \anhalf - \half)} 
	+ \frac{- \anhalf}{l(l - \anhalf)}
	- \frac{\anhalf}{l(l + \anhalf)}. \\
	\end{array}
}\hspace{-3em}
\end{equation}
The digamma function, denoted by $\psi(z)$, possesses a series expansion given by \cite[Eq. 6.3.16]{abramowitz_handbook_1964}
\begin{equation}
\begin{array}{l}
\psi(1 + z) = \shortminus\gamma + \hspace{-.5em}\sum\limits_{n=1}^\infty \frac{z}{n(n+z)}~\text{for}~z \notin \{\shortminus1,\shortminus2,\shortminus3,\dotsc\},
\end{array}
\end{equation}
where $\gamma$ denotes the Euler-Mascheroni constant.
We can thus simplify \ref{eq:unevenone} by expressing it in terms of the digamma function as
\begin{equation}
\label{eq:sumuneven}
\begin{array}{ll}
\sum\limits_{l=- \infty}^{\infty} \frac{(-1)^{l}}{\frac{\dcycfreq}{\blocksize} + l} &=
\frac{\blocksize}{\dcycfreq} + \frac{1}{2} \left(
\psi\left( \half + \anhalf \right) - \psi\left( \half - \anhalf \right) \right.\\
&\hspace{1cm} \left. + \psi\left( 1 - \anhalf \right) - \psi\left( 1 + \anhalf \right)
\right).
\end{array}
\end{equation}
Since the reflection and the recurrence formulas of the digamma function are known to be \cite[Eq. 6.3.7]{abramowitz_handbook_1964}
\begin{equation}
\begin{array}{l}
\psi(1-z) - \psi(z) = \pi \operatorname{cot}(\pi z)
\end{array}
\label{eq:recur}
\end{equation}
and \cite[Eq. 6.3.5]{abramowitz_handbook_1964}
\begin{equation}
\begin{array}{l}
\psi(1 + z) = \psi(z) + \frac{1}{z},
\end{array}
\label{eq:reflect}
\end{equation}
respectively, 
we obtain
\begin{equation}
\label{eq:digamma12}
\begin{array}{l}
\psi\left( \half + \anhalf \right) - \psi\left( \half - \anhalf \right) \\
= \psi\left( 1 - \left( \half - \anhalf \right) \right) - \psi\left( \half - \anhalf \right) \\
= \pi \operatorname{cot}\left(\pi \left(\half - \anhalf\right)\right)
\end{array}
\end{equation}
and
\begin{equation}
\label{eq:digamma34}
\begin{array}{l}
\psi\left( 1 - \anhalf \right) - \psi\left( 1 + \anhalf \right) \\
= \psi\left( 1 - \anhalf \right) - \psi\left( \anhalf \right) - 2 \frac{\blocksize}{\dcycfreq}\\
= \pi \operatorname{cot}(\pi \anhalf) - 2 \frac{\blocksize}{\dcycfreq}.
\end{array}
\end{equation}
Inserting  \ref{eq:digamma12} and \ref{eq:digamma34} into \ref{eq:sumuneven} results in
\begin{equation}
\label{eq:nsunevenfinal}
\begin{array}{l}
\sum\limits_{l=- \infty}^{\infty} \frac{(-1)^{l}}{\frac{\dcycfreq}{\blocksize} + l} 
= \half \pi \left( \operatorname{cot}\left(\pi \left( \half - \anhalf \right)\right) + \operatorname{cot}\left( \pi \anhalf \right) \right)\\
= \half \pi \left( \operatorname{tan}\left(\pi \anhalf \right) + \operatorname{cot}\left( \pi \anhalf \right) \right) = \frac{\pi}{ \operatorname{sin}\left( \pi \frac{\dcycfreq}{\blocksize} \right)}.
\end{array}
\end{equation}
Finally, substituting \ref{eq:nsunevenfinal} into \ref{eq:fullydiscreteca2} gives us the expression \ref{eq:finalanalca}.


\begin{backmatter}


\section*{Funding}
This work was partly supported by the Deutsche Forschungsgemeinschaft (DFG) projects CoCoSa (grant MA 1184/26-1) and CLASS (grant MA 1184/23-1).

\section*{Competing interests}
  The authors declare that they have no competing interests.


\bibliographystyle{bmc-mathphys} 
\bibliography{cacafe_paper}      




\begin{figure}[h!]
	\includegraphics{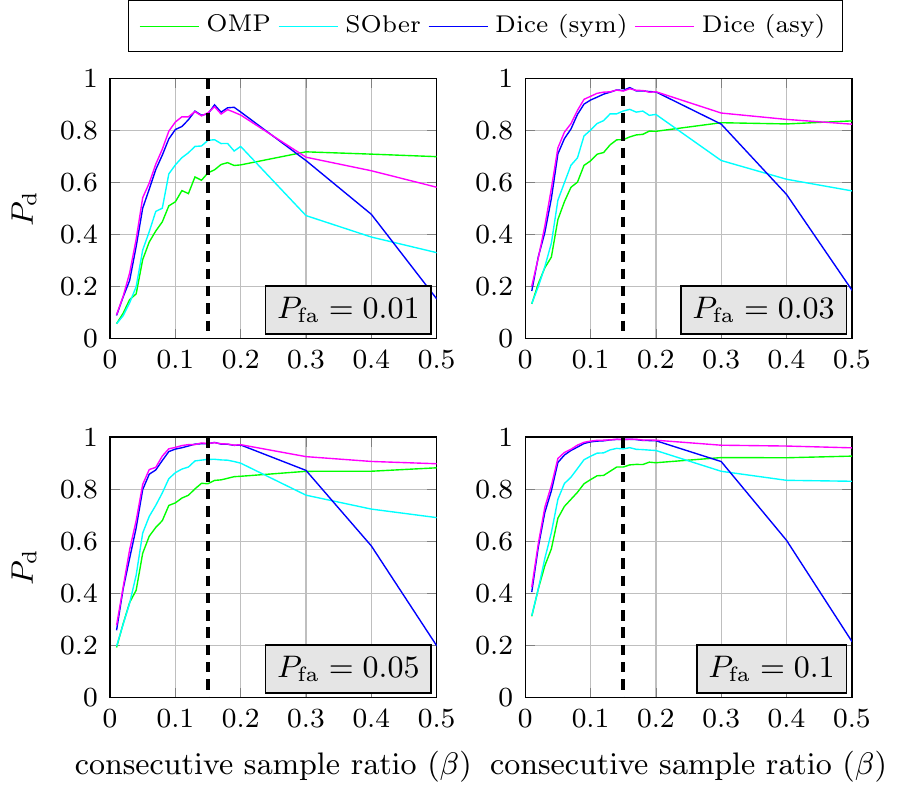}
	\caption{Detection rate over consecutive sample ratio for different false alarm rates at $0$ dB \ac{snr}.
	}
	\label{fig:pd_over_consec}
\end{figure}

\begin{figure}[h!]
	\includegraphics{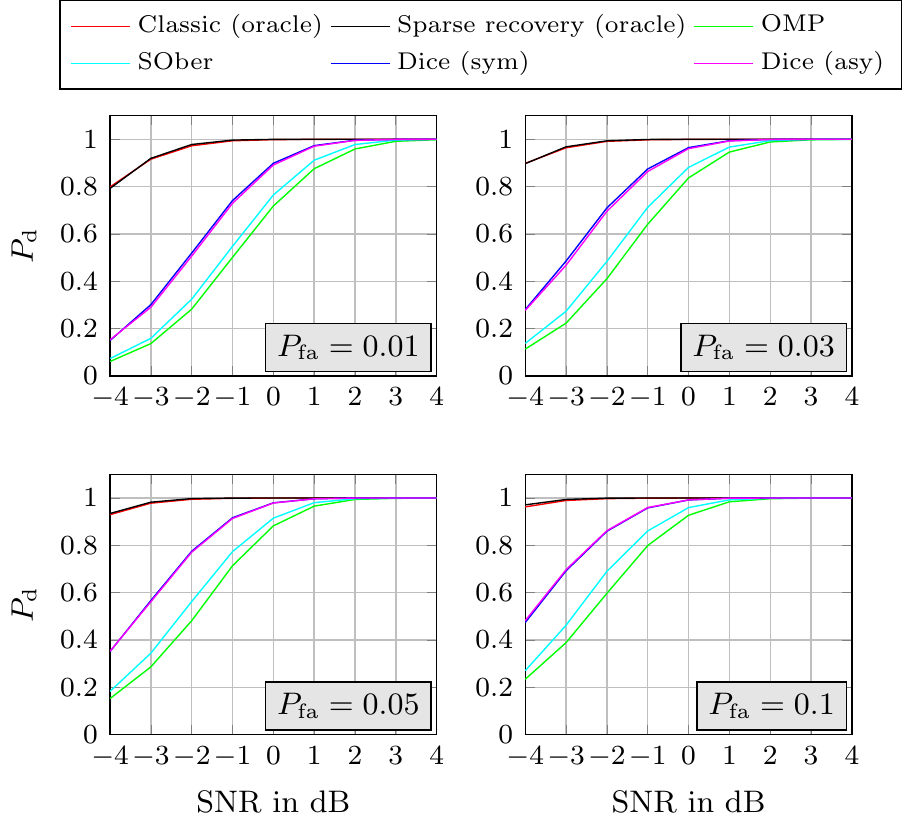}
	\caption{Maximum detection rate (optimal individual consecutive sample ratio selection) over \ac{snr} for different false alarm rates.
	}
	\label{fig:max_pd_over_snr}
\end{figure}

\begin{figure}[h!]
	\includegraphics{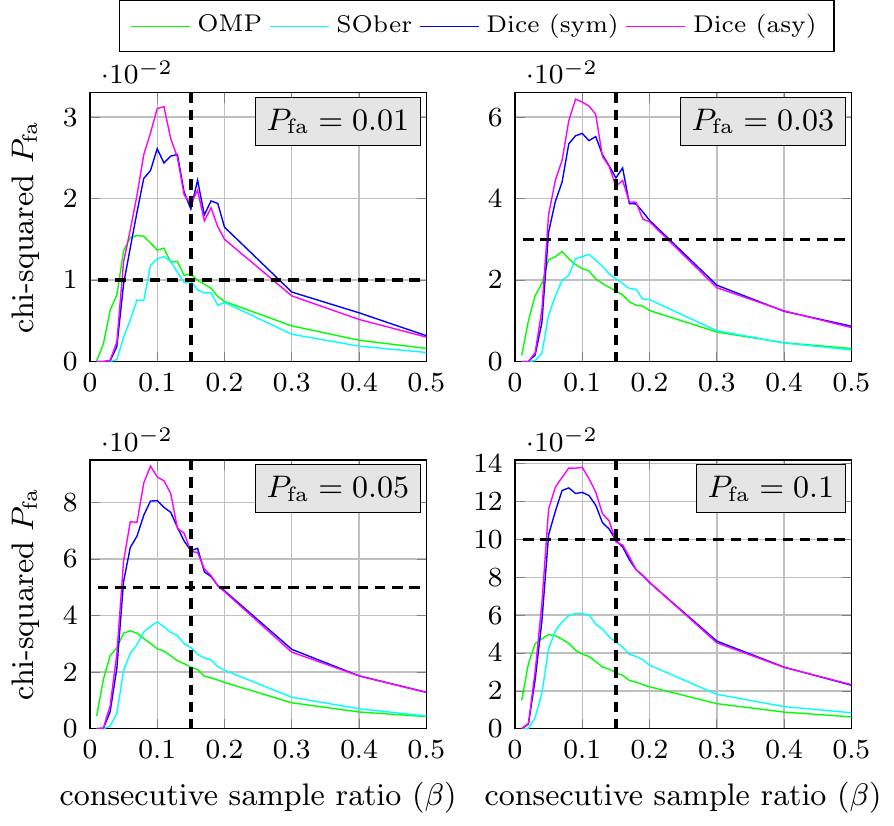}
	\caption{False alarm rate that has to be selected according to the chi-squared distribution to obtain different actual false alarm rates over the consecutive sample ratio.
	}
	\label{fig:far_settings}
\end{figure}

\begin{figure}[h!]
	\includegraphics{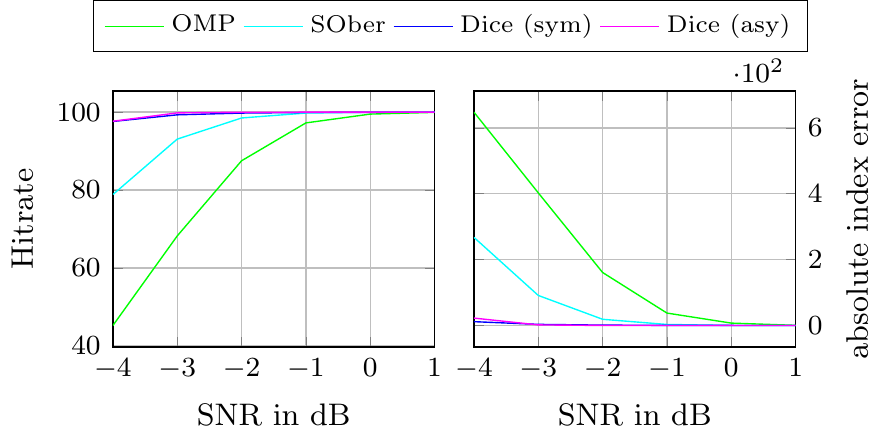}
	\caption{Left: Hitrate over \ac{snr}, right: absolute index error over \ac{snr}. Both at consecutive sample ratio 0.15.
	}
	\label{fig:hitrates_absidxerrors}
\end{figure}

\begin{figure}[h!]
	\includegraphics{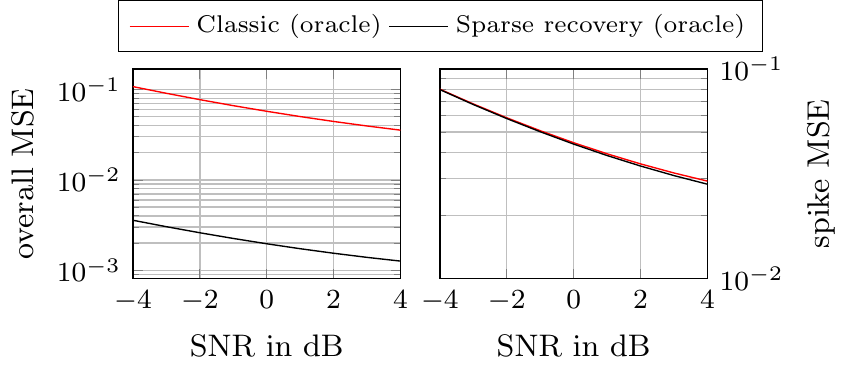}
	\caption{MSE between the CAF estimation and the actual (analytic) value. Left: over the whole support, right: at the cycle frequencies. Both at consecutive sample ratio 0.15.
	}
	\label{fig:caf_mse}
\end{figure}


\begin{table}[h!]
	\caption{System Parameters
	}
	\label{tab:sim_par}
	\centering
	{
		\begin{tabular}{l c c}
			\hline 
			Parameter & Symbol & Value(s) \\
			\hline 
			Size of the CA vector - \ac{cs} methods & $\blocksize$ & $4000$ \\
			\# of known delay-product elements & $\navailsmpls$ & $1000$ \\
			Size of the CA vector - classic method & $\navailsmpls$ & $1000$ \\
			Discrete time delays & $\ddelay$ & $\{1, 2, 3, 4\}$ \\
			Modulation type & & BPSK\\
			Discrete symbol length & $\nsym$ & $8$\\
			Signal to noise ratio & \ac{snr} & $\{-4, \dotsc, 4\}$dB\\
			\# of Monte Carlo instances & & 10000\\
			Consecutive sample ratio & $\consecfrac$ & $\{0.01, \dotsc, 0.5 \}$\\
			Covariance estimation window type & $W$ & Kaiser\\
			Kaiser window parameter & $\alpha_\text{Kaiser}$ & $10$\\
			Kaiser window length & $L$ & $201$\\
			\hline
		\end{tabular}
	}
\end{table}

\end{backmatter}
\end{document}